# Controlled interfacial assembly of 2D curved colloidal crystals and jammed shells

Anand Bala Subramaniam, Manouk Abkarian and Howard A. Stone

Division of Engineering and Applied Sciences, Harvard University, Cambridge, MA 02138

**Assembly of colloidal particles on fluid interfaces is a promising technique for synthesizing two-dimensional micro-crystalline materials useful in fields as diverse as biomedicine[1], materials science[2], mineral flotation[3] and food processing[4]. Current approaches rely on bulk emulsification methods, require further chemical and thermal treatments, and are restrictive with respect to the materials employed[5-9]. The development of methods that exploit the great potential of interfacial assembly for producing tailored materials have been hampered by the lack of understanding of the assembly process. Here we report a microfluidic method that allows direct visualization and understanding of the dynamics of colloidal crystal growth on curved interfaces. The crystals are periodically ejected to form stable jammed shells, which we refer to as colloidal armour. We propose that the energetic barriers to interfacial crystal growth and organization can be overcome by targeted delivery of colloidal particles through hydrodynamic flows. Our method allows an unprecedented degree of control over armour composition, size and stability.**

We designed a three-channel hydrodynamic focusing device in order to control particle delivery and assembly at the scale of individual shells (Fig. 1a). A suspension of colloidal particles is driven through the outer channels, while the inner channel carries the dispersed phase (gas or liquid). The interface between the continuous and dispersed phases serves as the substrate for crystal assembly. In the example shown in Figs. 1b,c, 4 µm diameter charge-stabilized fluorescent polystyrene beads (the large size is chosen to facilitate visualization) assemble as a monolayer on a gas thread, which eventually becomes unstable, and breaks to form a spherical shell (Movie 1).

The curved interface is held stationary relative to the motion of the particles in the continuous phase, to allow sufficient time for the particles to adsorb and produce a close-packed crystallite (Details provided in methods section). Typically, for a particle concentration of 0.1 volume percent, we found that armoured bubbles are ejected at a rate of 10 per second. Higher formation frequencies can be obtained by using a more concentrated suspension and tuning the driving pressures of the two phases. The shear-driven ejection of the jammed shells makes the geometry of the outlet channel the main determinant of the size of the armoured object, and the continuous one-step assembly process produces monodisperse armoured gas bubbles (Fig. 1d) (Movie 2). A close-up of the armoured bubble reveals the tightly-packed jammed structure of the colloidal shell (Fig. 1e), a feature which confers intrinsic stability as we discuss in more detail below.

We observed that jammed shells are only reproducibly shed at high particle velocities (10 cm/s), which suggests that each capture event (i.e. where a particle is adsorbed at the interface) occurs at a timescale of tens of microseconds. We took high-speed video at 70,000 frames per second to better understand the dynamics of the assembly. We found that the colloidal particles have a well defined region of capture, characterized by the largest change in flow lines, the minima indicated by the solid line in Figs. 2a, d. Particles that approach the interface at or above this line are captured when they collide with the interface (Fig. 2b). The particles are trapped on the interface in a deep potential energy well[10], estimated to be $10^7$ kT for micron diameter particles, and we observe that the shear experienced by the particles from the surrounding fluid is insufficient to detach adsorbed particles. Instead the particles are transported by the flow to the anterior of the curved interface (Movie 3). Repeated capture and transport events result in a rapid build up of particles in a tightly packed crystallite, with the particles stacking in parallel lines closely following the flow of the continuous phase (Fig. 2c).

The geometry of the focusing device ensures that particle capture occurs far from the growing armour, which avoids steric and electrostatic repulsions[11] a particle would otherwise experience if it directly approaches a growing shell. This geometric advantage allows the formation of more densely packed crystallites than possible through current methods of emulsion templating[5-9]. Occasional missing particles along a stack cause rearrangements through the tumbling of adjacent particles, which thus self-heals most defects in the interfacial packing. Fig. 3a shows a single armoured bubble revealing the well-ordered structure of the colloidal armour. Triangulation of the particles, shown in Fig. 3b, reveals paired 5-7 defects in a six-fold coordinated lattice consistent with theoretical predictions of ground states of spherical crystals[12].

However, unlike thermally equilibrated spherical crystals, the energetically costly dislocations do not diffuse on the interface[13], revealing a freezing of the dynamics of the shell. Moreover, for colloidal armour composed of Brownian particles, such as 1 μm polystyrene beads, thermal motion apparent in the bulk phase is arrested in the tightly packed shell. This observation is in contrast with previously reported particle shells, where Brownian motion is often observed on the droplet interface[5,6,14]. It has been postulated recently that jammed particulate systems undergo a universal liquid-to-solid transition characterized by a reduced set of parameters, i.e. applied stress or pressure, particle number density, and the energy of interactions in the system[15,16]. We point out that in the unique case of particle-covered droplets (gas or liquid), surface tension provides a two-dimensional isotropic compressive stress and the finite spherical topology of the droplet results in unbounded confinement of the interfacially trapped particles. Such confinement is required for a jamming transition of repulsive particles[16], such as for the charged stabilized colloids employed here (confinement is clearly not necessary for attractive particles).

We thus believe that the high particle number densities obtained through our method of targeted delivery results in the spontaneous jamming of the particles in the shell. Such jammed systems undergo shear induced liquid-to-solid transitions, which confers great stability to the droplet: the jammed shell not only resists spontaneous coalescence due to interfacial surface area minimization[17], it should also be able to resist shear-induced coalescence. Based on our many experiments with different materials for the continuous and dispersed phases and for the particles, this topologically induced jamming appears to be independent of the nature of the particle-particle interactions. The jammed armour is thus intrinsically stable, and spontaneous coalescence of the fully armoured droplets with each other or with an uncovered interface is never observed (Fig. 1d).

Visualization of the process of particle capture and assembly leads us to believe that specific input of energy is required for the assembly of close-packed interfacial crystals. In the absence of flow, we never observe spontaneous adsorption of micron-size particles onto emulsion droplets to form particle shells. Addition of NaCl, up to a final concentration of 1.0 M, which reduces any particle/interface electrostatic repulsion[11], resulted in aggregation in the bulk of the suspension, but the aggregates did not adsorb and cover the droplets, within the timescale of a 10 minute experiment. Gentle agitation of the suspension was performed to prevent particle sedimentation, but coverage of the interface was still not achieved. More forceful agitation however produces partially covered droplets as observed for particle-stabilized emulsions[14]. We thus propose that production of colloidal shells on fluid droplets requires specific input of energy through hydrodynamic flows to overcome barriers to adsorption[11,18,19]. Indeed, the particles' non-dimensional Reynolds number in our device, $Re_p = Ua/\nu$ is estimated to be 0.2, where $U$ is the particle speed, $a$ is the particle radius, and $\nu$ is the kinematic viscosity of water, reflecting the non-negligible role of inertia on the capture events. We note that the production of colloidal shells described in the literature, which has been characterized as a self-assembled system[7,14], was always in batch concomitant with the necessary emulsification steps. The complicated hydrodynamic flows and mixing produced during typical emulsification processes results in shell formation, which from a post-synthesis vantage point may appear to be a self-assembly process.

The generality of our concept and demonstration of flow-driven assembly is further exemplified by the wide range of armour/core objects we produce without surface modification of the colloidal particles. Typically, hydrophilic particles such as silica spheres suspended in water are chemically modified to increase hydrophobicity before adsorption is observed at the air/water interface or at the interface of oil droplets[3,5,6,20]. Such surface modification restricts particle/core combinations to intermediate wetting and, moreover, reduces the efficiency of synthesis, since the hydrophobic particles flocculate rapidly in the bulk aqueous phase[21]. An alternative means of stabilizing particles at the interface is to increase the dimensions of the particle. It is to be expected that the energetic barrier to adsorption increases with increased total charge, i.e. larger particles (assuming constant surface charge density)[3]. This increased energetic cost of forcing micron-size particles onto an interface, however, is easily provided by our targeted hydrodynamic flows. For example, we have manufactured air and oil droplets armoured with unmodified silica particles initially dispersed in water (Fig. 4a). Shells composed of micron-size hydrophilic silica and gold particles are as stable as those produced from hydrophobic polystyrene, and similar arrest of Brownian motion in the jammed shells is observed. Thus, shell/core combinations, such as a fully hydrophobic shell around a hydrophilic core and vice versa, can be produced. We have also manufactured armour with unmodified conductive metallic particles, a feature, as far as we know, unreported previously (Fig. 4b). Armour with various dielectric properties can be obtained by varying the conductivity of the particles or the ratio of conductive and insulating particles on the interface; in the case of a fully conductive armour the fluid core should then be protected from stray electromagnetic fields (analogous to a Faraday cage).

An appreciation of the fundamental energetic principles of interfacial assembly allows us to go a step further and tailor new kinds of shells, with precise combinations and relative positions of particles on the interface. By loading particles differentially labelled with rhodamine and fluorescein in the two outer channels of the microfluidic device, we were able to produce hemi-shells, or Janus armour (Fig. 5). The ability to assemble two or more types of particles on a single shell is a first step in producing chemically patterned shells that may be useful for targeting or sorting purposes.

In conclusion, our method of controlled flow-driven assembly overcomes the energetic barriers to interfacial crystal growth to produce jammed colloidal shells. The jammed state confers the armour intrinsic stability, thus negating the need for external locking steps. Furthermore, the generality of our method suggests new applications and many strategies for making advanced patterned colloidal shells of controlled size, shape, and composition.

**Figures:**

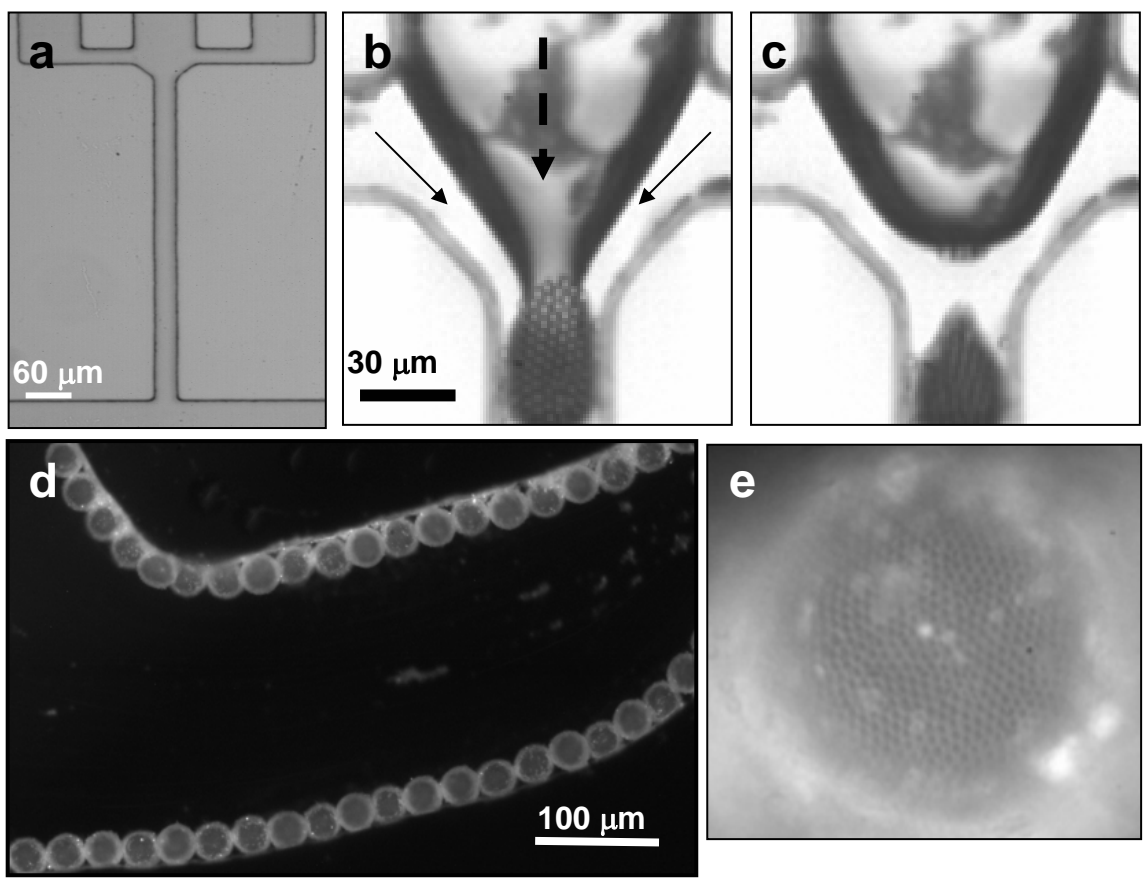

**Figure 1: Hydrodynamic focusing for the continuous formation of droplets with colloidal armour**. **a,** Picture of the flow-focusing device for flow-driven assembly of colloidal armour. Fluids are driven through the channels by connecting custom syringes to a supply of compressed air. The dispersed phase is injected through the inner channel, while the particle containing continuous phase is injected through the outer channels. A 1mm X 1mm view chamber is constructed at the end of the device. **b,** The direction of dispersed phase flow is indicated by a dashed arrow, while the direction of the colloidal suspension flow is indicated by the solid arrows. Flow of the continuous phase liquid focuses the dispersed phase fluid into a narrow thread and targets the particles onto the interface. The speed of the particles and the frequency of shedding can be controlled by tuning the difference in driving pressures of the inner and outer channels (typical values

are suspension: 1.92 psi, gas: 1.69 psi). The interfacial crystal consisting of 4 µm diameter charge-stabilized fluorescent polystyrene beads grows and subsequently experiences greater shear **c,** which results in the ejection of a jammed shell. **d,** Monodisperse stable armoured bubbles arrayed along a gas/water interface in the view chamber. The apparent minor size differential is due to the differing heights of the shells from the focal plane of the objective. The armoured bubbles do not coalesce even when closely apposed and seem to be stable indefinitely. **e,** A close-up of a single bubble shown in **d,** the Brownian particles (1.0 µm in diameter) are jammed in position.

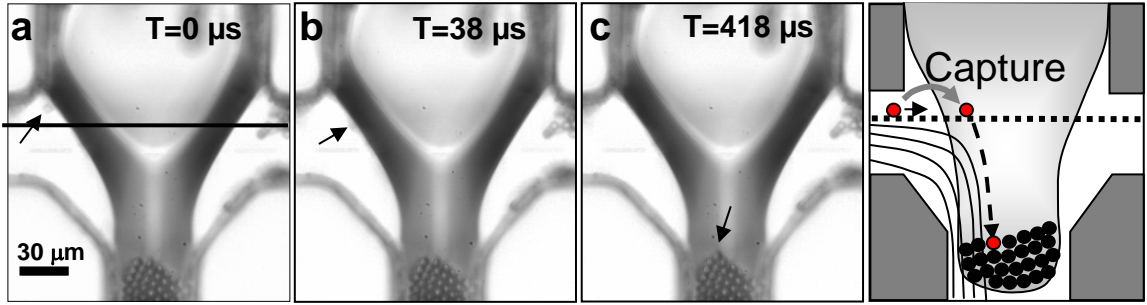

**Figure 2: Time-lapse sequence of the capture of an individual particle (particle indicated with a black arrow) a,** Particles are captured when they approach the interface, at or above the indicated flow line. **b,** The particle approaches the interface at a speed of 10 cm/s and adsorbs on the interface far from the other adsorbed particles. **c,** The shear experienced by the particle from the flow, drives it to a stagnation point at the anterior of the curved interface. Particle movement is stopped by previously stacked particles. Particles stack on the interface following the flow lines, with most of the particles arranged in a six-fold coordinated lattice. **d,** Schematic of the capture process. Particles that approach the interface above the solid line, undergo a large change in flow trajectory as the fluid enters the outlet channel, forcing the adsorption of the particle on the interface. Particles on flow lines below the solid line undergo smaller changes in trajectory and do not posses sufficient energy to overcome barriers to adsorption.

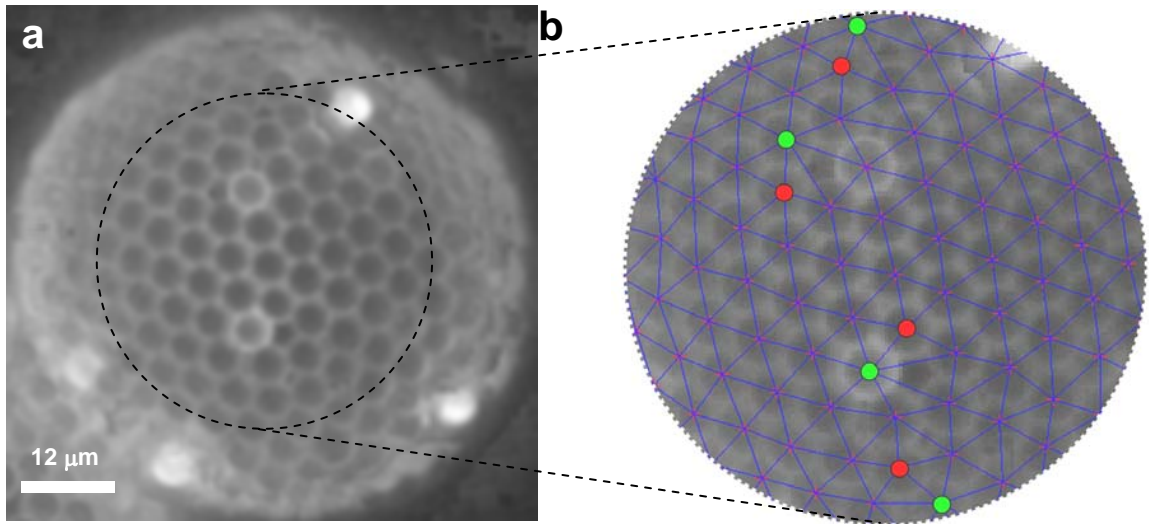

**Figure 3: Packing on the jammed shell. a,** Close-up of a single armoured bubble, revealing the close-packed nature of the rigid shell. **b,** Triangulation of the shell reveals several 5-7 paired dislocations on a six-fold coordinated lattice consistent with theoretical predictions of ground states of spherical crystals[12].

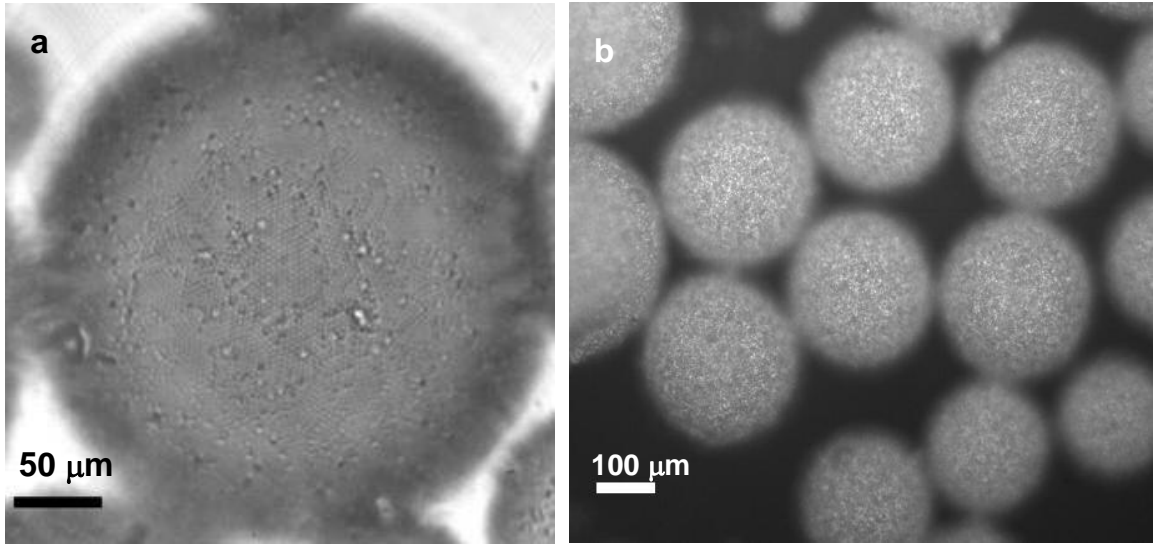

**Figure 4: Various shell/core combinations**. **a,** Unmodified silica particles on a mineral oil droplet. This is an example of colloidal armor composed of hydrophilic particles from a hydrophilic continuous phase adsorbed on a hydrophobic droplet. **b,** Polydisperse gold particles on gas bubbles dispersed in water. This is an example of conductive armor on a gas bubble. The polydispersity of the particles employed resulted in a slight dispersion in the size of the jammed shells produced.

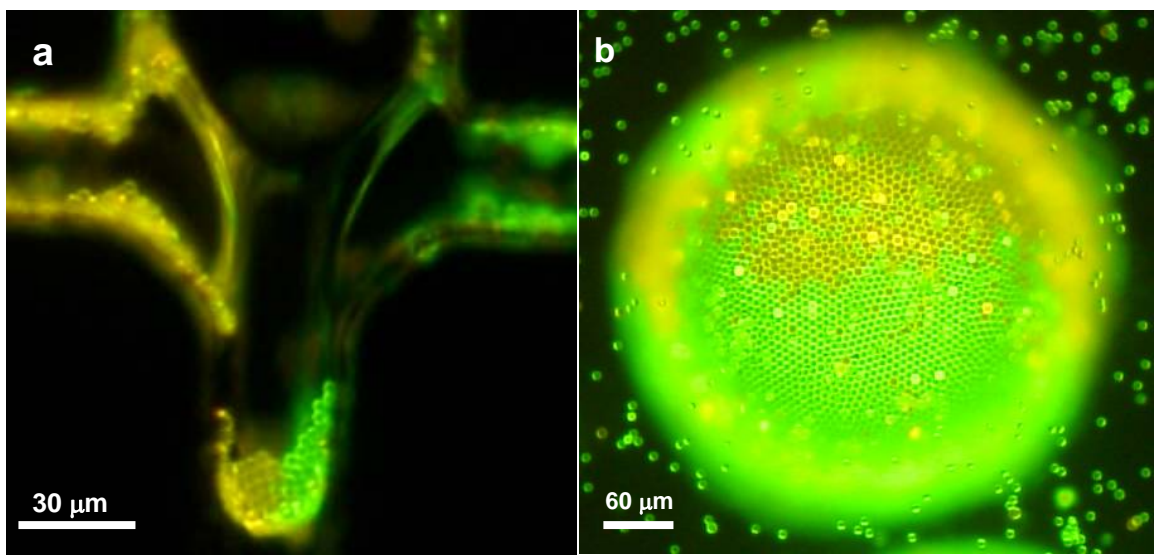

**Figure 5: Tailored production of Janus armour a,** Assembly of particles on an air/water interface to produce Janus crystals. The yellow particles are 4.9 μm diameter polystyrene particles dyed with rhodamine, while the green particles are 4.0 μm particles dyed with fluorescein. **b,** An example of the Janus shell, with approximately two hemispheres of different size particles and fluorescence.

We thank R. Larsen, A. Lips, L. Mahadevan, and R. Subramanian for helpful conversations. Support from Unilever Research and the Harvard Material Research Science and Engineering Center (DMR-0213805) are gratefully acknowledged.

**Methods:**

The microfluidic device was manufactured using principles of soft-lithography introduced by Whitesides and coworkers.

Surfactant-free fluorescent polystyrene particles (IDC Corporation) were diluted with purified water (Millipore) to a typical concentration of less than 0.1 volume percent. The particles are described as being hydrophobic with a small amount of surface charge providing colloidal stability in suspension. The colloidal suspension is filled in a gas tight syringe (Hamilton) and connected to a compressed gas tank through custom adapters. Polyethylene tubes are connected from the syringe to the continuous phase inlet hole of the device. A similar syringe without the suspension was connected to the dispersed phase inlet hole of the device. Pressure applied to the suspension in the syringe was independently controlled by a regulator (Bellofram) with a precision of 0.001 psi. We found that driving the fluids using an applied pressure, allowed more rapid control over the flow velocities than traditional flow-rate controlled syringe pumps.

To ensure reproducible production of fully armored objects, we worked in a pressure regime where the difference of driving pressures between the dispersed and continuous phase was small, on the order of 0.2-0.5 psi. For larger differences in driving pressures, incomplete coverage of the gas bubbles is observed, and the gas bubbles are unstable to coalescence in the view chamber. The dispersed phase used was varied. We made droplets with gaseous argon, $CO_2$, $O_2$, $N_2$, and liquid mineral oil (Sigma) and octanol (Sigma). The colloidal particles were also varied. We used monodisperse polystyrene particles of 1.6 μm, 2.1 μm, 4.0 μm, and 4.6 μm diameter, 1.6 μm diameter silica particles (Bangs Lab), 1.0 μm diameter PMMA particles (Bangs Lab), and polydisperse agglomerated gold microparticles, with mean diameters ranging from 1.0-3.99 μm (Sigma). All the particles were diluted with ultrapure water to typical concentrations of 0.1 volume percent.